\documentclass[sigconf]{acmart}

\usepackage{booktabs} 
\usepackage{multirow}
\usepackage{mathrsfs}
\usepackage{amsmath}
\usepackage{balance}

\copyrightyear{2018}
\acmYear{2018} 
\setcopyright{iw3c2w3}
\acmConference[WWW 2018]{The 2018 Web Conference}{April 23--27, 2018}{Lyon, France}
\acmBooktitle{WWW 2018: The 2018 Web Conference, April 23--27, 2018, Lyon, France}
\acmPrice{}
\acmDOI{10.1145/3178876.3186030}
\acmISBN{978-1-4503-5639-8/18/04}

\begin{document}
\title{CESI: Canonicalizing Open Knowledge Bases using \\ Embeddings and Side Information}

\author{Shikhar Vashishth}
\orcid{0000-0002-6258-2494}
\affiliation{
	\institution{Indian Institute of Science}
	\city{Bangalore} 
	\state{India}
}
\email{shikhar@iisc.ac.in}

\author{Prince Jain}\authornote{Research carried out while at the Indian Institute of Science, Bangalore.}
\affiliation{
	\institution{Microsoft}
	\city{Bangalore} 
	\state{India}
}
\email{prince.jain@microsoft.com}

\author{Partha Talukdar}
\affiliation{
	\institution{Indian Institute of Science}
	\city{Bangalore} 
	\state{India}
}
\email{ppt@iisc.ac.in}

\renewcommand{\shortauthors}{S. Vashishth et al.}

\newcommand\BibTeX{B{\sc ib}\TeX}

\newcommand{\refalg}[1]{Algorithm~\ref{#1}}
\newcommand{\refeqn}[1]{Equation~\ref{#1}}
\newcommand{\reffig}[1]{Figure~\ref{#1}}
\newcommand{\reftbl}[1]{Table~\ref{#1}}
\newcommand{\refsec}[1]{Section~\ref{#1}}
\newcommand{\method}{CESI}

\newcommand{\methodfull}{Canonicalization using Embeddings and Side Information}

\newcommand{\gold}{L}
\newcommand{\elink}[2]{\mathcal{E}_{#1}(#2)}
\newcommand{\rlink}[2]{\mathcal{R}_{#1}(#2)}

\definecolor{mycolor}{rgb}{0,0.4,0}
\newcommand{\sv}[1]{\textcolor{mycolor}{#1}}

\newcommand*{\Comb}[2]{{}^{#1}C_{#2}}

\newcommand{\NP}{NP}
\newcommand{\RP}{relation phrase}
\newcommand{\sideConst}[2]{\lambda_{\text{#1}, #2}}
\newcommand{\sidePairs}[2]{\mathcal{Z}_{\text{#1}, #2}}

\newcommand{\stepRep}{Source based NP segregation}
\newcommand{\stepSide}{Side Information Acquisition}
\newcommand{\stepEmbed}{Embedding NP and Relation Phrases}
\newcommand{\stepCluster}{Clustering Embeddings and Canonicalization}
\newcommand{\stepCanon}{Selecting Canonical Representative from Cluster}

\newcommand{\myData}{ReVerb45K}

\newcommand{\gidf}{Gal\'arraga-IDF}
\newcommand{\gstrsim}{Gal\'arraga-StrSim}
\newcommand{\gattr}{Gal\'arraga-Attr}

\newcommand{\note}[1]{}

\begin{abstract}
Open Information Extraction (OpenIE) methods extract \textit{(noun phrase, relation phrase, noun phrase)} triples from text, resulting in the construction of large Open Knowledge Bases (Open KBs). The noun phrases (NPs) and relation phrases in such Open KBs are not \emph{canonicalized}, leading to the storage of redundant and ambiguous facts. Recent research has posed canonicalization of Open KBs as clustering over \emph{manually-defined} feature spaces. Manual feature engineering is expensive and often sub-optimal. In order to overcome this challenge, we propose Canonicalization using Embeddings and Side Information (\method{}) -- a novel approach which performs canonicalization over \emph{learned} embeddings of Open KBs. \method{} extends recent advances in KB embedding by incorporating relevant NP and \RP{} side information in a principled manner. Through extensive experiments on multiple real-world datasets, we demonstrate \method{}'s effectiveness. 
\end{abstract}

%
%



\maketitle

\section{Introduction}
\label{sec:intro}

Recent research has resulted in the development of several large \emph{Ontological} Knowledge Bases (KBs), examples include  DBpedia \cite{Auer:2007:DNW:1785162.1785216}, YAGO \cite{Suchanek:2007:YCS:1242572.1242667}, and Freebase \cite{Bollacker:2008:FCC:1376616.1376746}. These KBs are called  ontological as the knowledge captured by them conform to a fixed ontology, i.e., pre-specified Categories (e.g., \textit{person, city}) and Relations (e.g., \textit{mayorOfCity(Person, City)}). Construction of such ontological KBs require significant human supervision. Moreover, due to the need for pre-specification of the ontology, such KB construction methods can't be quickly adapted to new domains and corpora. While other ontological KB construction approaches such as NELL \cite{NELL-aaai15} learn from limited human supervision, they  still suffers from the quick adaptation bottleneck.

In contrast, Open Information Extraction (OpenIE) methods need neither supervision nor any pre-specified  ontology. Given unstructured text documents, OpenIE methods readily extract triples of the form \textit{(noun phrase, relation phrase, noun phrase)} from them, resulting in the development of large Open Knowledge Bases (Open KBs). Examples of Open KBs include TextRunner \cite{Banko:2007:OIE:1625275.1625705}, ReVerb \cite{Fader:2011:IRO:2145432.2145596}, and OLLIE \cite{ollie1,ollie2,ollie3}. While this makes OpenIE methods highly adaptable, they suffer from the following shortcoming: unlike Ontological KBs, the Noun Phrases (NPs) and \RP{}s in Open KBs are not \emph{canonicalized}. This results in storage of redundant and ambiguous facts.


Let us explain the need for canonicalization through a concrete example. Please consider the two sentences below.

\begin{center}
\textit{Barack Obama was the president of US.} \\
\textit{Obama was born in Honolulu.}
\end{center}

Given the two sentences above, an OpenIE method may extract the two triples below and store them in an Open KB.

\begin{center}
\textit{(Barack Obama, was president of, US)} \\
\textit{(Obama, born in, Honolulu)}
\end{center}

Unfortunately, neither such OpenIE methods nor the associated Open KBs have any knowledge that both \textit{Barack Obama} and \textit{Obama} refer to the same person. This can be a significant problem as Open KBs will not return all the facts associated with \textit{Barack Obama} on querying for it. Such KBs will also contain redundant facts, which is undesirable. Thus, there is an urgent need to \emph{canonicalize} noun phrases (NPs) and relations in Open KBs.

In spite of its importance, canonicalization of Open KBs is a relatively unexplored problem. In \cite{Galarraga:2014:COK:2661829.2662073}, canonicalization of Open KBs is posed as a clustering problem over \emph{manually} defined feature representations. Given the costs and sub-optimality involved with manual feature engineering, and inspired by recent advances in knowledge base embedding \cite{NIPS2013_5071,Nickel:2016:HEK:3016100.3016172}, we pose canonicalization of Open KBs as a clustering over \emph{automatically learned} embeddings. We make the following contributions in this paper.
\begin{itemize}
	\item We propose Canonicalization using Embeddings and Side Information (\method{}), a novel method for canonicalizing Open KBs using learned embeddings. To the best of our knowledge,
	this is the first approach to use learned embeddings and side information
	for canonicalizing an Open KB.
	\item {\method} models the problem of noun phrase (NP) and relation phrase canonicalization \emph{jointly} using relevant side information  
	in a principled manner. This is unlike  prior approaches where \NP{} and \RP{} canonicalization were performed sequentially.
	\item We build and experiment with \myData{}, a new dataset for Open KB canonicalization. \myData{} consists of 20x more NPs than the previous biggest dataset for this task. Through extensive experiments on this and other real-world datasets, we demonstrate CESI's  effectiveness (\refsec{sec:experiments}).
\end{itemize}

\method{}'s source code and datasets used in the paper are available at \texttt{\url{https://github.com/malllabiisc/cesi}}.

\section{Related Work}
 \label{sec:related}

{\bf Entity Linking}: One traditional approach to canonicalizing noun phrases is to map them to an existing KB such as Wikipedia or Freebase. This problem is known as Entity Linking (EL) or Named Entity Disambiguation (NED). Most approaches generate a list of candidate entities for each NP and re-rank them using machine learning techniques. Entity linking has been an active area of research in the NLP community \cite{Trani:2014:DOS:2878453.2878558,Lin:2012:ELW:2391200.2391216,Ratinov:2011:LGA:2002472.2002642}. A major problem with these kind of approaches is that many NPs may refer to new and emerging entities which may not exist in KBs. One approach to resolve these noun phrases is to map them to NIL or an OOKB (Out of Knowledge Base) entity, but the problem still remains as to how to cluster these NIL mentions. Although entity linking is not the best approach to NP canonicalization, we still leverage signals from entity linking systems for improved canonicalization in \method{}.

{\bf Canonicalization in Ontological KBs}: Concept Resolver \cite{Krishnamurthy:2011:NPD:2002472.2002545} is used for clustering NP mentions in NELL \cite{NELL-aaai15}. It makes ``one sense per category" assumption which states that a noun phrase can refer to at most one concept in each category of NELL's ontology. For example, the noun phrase ``Apple" can either refer to a company or a fruit, but it can refer to only one company and only one fruit. Another related problem to NP canonicalization is Knowledge Graph Identification \cite{Pujara:2013:KGI:2717129.2717164}, where given a noisy extraction graph, the task is to produce a consistent Knowledge Graph (KG) by performing entity resolution, entity classification and link prediction jointly. Pujara et al. \cite{Pujara:2013:KGI:2717129.2717164} incorporate information from multiple extraction sources and use ontological information to infer the most probable knowledge graph using probabilistic soft logic (PSL) \cite{Brocheler:2010:PSL:3023549.3023558}. However, both of these approaches require additional information in the form of an ontology of relations, which is not available in the Open KB setting.

{\bf Relation Taxonomy Induction}: SICTF \cite{D16-1040} tries to learn relation schemas for different OpenIE relations. It is built up on RESCAL \cite{Nickel:2011:TMC:3104482.3104584}, and uses tensor factorization methods to cluster noun phrases into \emph{categories} (such as ``person", ``disease", etc.). We, however, are interested in clustering noun phrases into entities.

There has been relatively less work on the task of relation phrase canonicalization. Some of the early works include DIRT \cite{Lin:2001:DSI:502512.502559}, which proposes an unsupervised method for discovering inference rules of the form ``\textit{X is the author of Y }  $\approx$ \textit{X wrote Y}" using paths in dependency trees; and the PATTY system \cite{Nakashole:2012:PTR:2390948.2391076}, which tries to learn subsumption rules among relations (such as \textit{son-of} $\subset$ \textit{child-of}) using techniques based on frequent itemset mining. These approaches are more focused on finding a taxonomy of relation phrases, while we are looking at finding equivalence between relation phrases.

{\bf Knowledge Base Embedding}: KB embedding techniques such as TransE \cite{NIPS2013_5071}, HolE \cite{Nickel:2016:HEK:3016100.3016172} try to learn vector space embeddings for entities and relations present in a KB. TransE makes the assumption that for any \textit{$\langle$subject, relation, object$\rangle$} triple, the relation vector is a translation from the subject vector to the object vector. HolE, on the other hand, uses non-linear operators to model a triple. These embedding methods have been successfully applied for the task of link prediction in KBs. In this work, we build up on HolE while exploiting relevant side information for the task of Open KB canonicalization. We note that, even though KB embedding techniques like HolE have been applied to ontological KBs, \method{} might be the first attempt to use them in the context of Open KBs.

{\bf Canonicalizing Open KBs}: The RESOLVER system \cite{Yates:2009:UMD:1622716.1622724} uses string similarity based features to cluster phrases in TextRunner \cite{Banko:2007:OIE:1625275.1625705} triples. String similarity features, although being effective, fail to handle synonymous phrases which have completely different surface forms, such as \textit{Myopia} and \textit{Near-sightedness}. 

KB-Unify \cite{dellibovi-espinosaanke-navigli:2015:EMNLP} addresses the problem of unifying multiple Ontological and Open KBs into one KB. However, KB-Unify requires a pre-determined sense inventory which is not available in the setting \method{} operates. 


The most closely related work to ours is \cite{Galarraga:2014:COK:2661829.2662073}. They perform NP canonicalization by performing Hierarchical Agglomerative Clustering (HAC) \cite{Tan:2005:IDM:1095618} over manually-defined feature spaces, and subsequently perform  relation phrase  clustering by using the AMIE algorithm \cite{Galarraga:2013:AAR:2488388.2488425}. \method{} significantly outperforms this prior method (\refsec{sec:experiments}). 

\section{Proposed Approach: \method{}}
\label{sec:approach}

\begin{figure*}
	\begin{center}
	\includegraphics[scale=0.5]{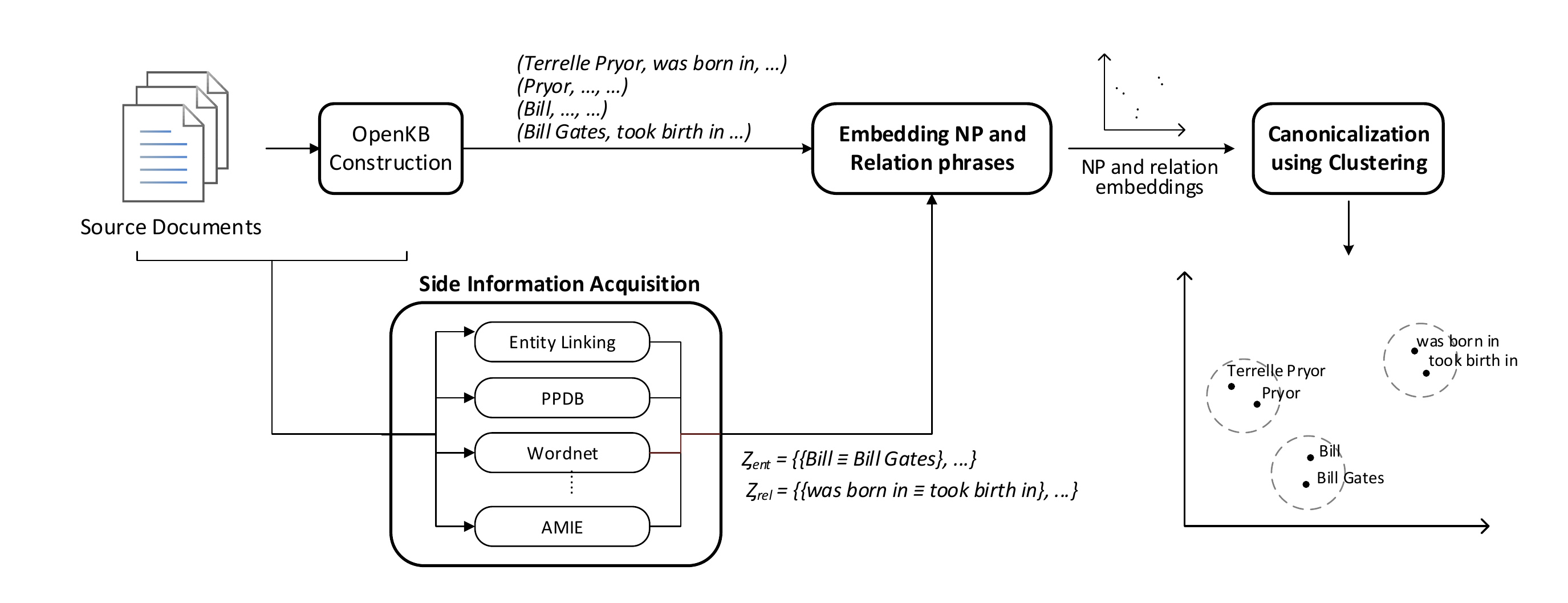}
	\caption{\label{fig:overview}Overview of \method{}. \method{} first acquires side information of noun and relation phrases of Open KB triples. In the second step, it learns embeddings of these NPs and relation phrases while utilizing the side information obtained in previous step. In the third step, \method{} performs clustering over the learned embeddings to canonicalize NP and relation phrases. Please see \refsec{sec:approach} for more details.}
	\end{center}
\end{figure*}

Overall architecture and dataflow of \method{} is shown in Figure \ref{fig:overview}. The input to \method{} is an un-canonicalized Open Knowledge Base (KB) with source information for each triple. The output is a list of canonicalized noun and relation phrases, which can be used to identify equivalent entities and relations or canonicalize the KB. \method{} achieves this through its three step procedure:\\\\

\begin{enumerate}
	\item \textbf{\stepSide:} The goal of this step is to gather various NP and \RP{} side information for each triple in the input by running several standard algorithms on the source text of the triples. More details can be found in Section \ref{sec:side-info}.
	\item \textbf{\stepEmbed:} In this step, CESI learns specialized vector embeddings for all NPs and \RP{}s in the input by making principled use of side information available from the previous step. 
	\item \textbf{\stepCluster:} Goal of this step is to cluster the NPs and \RP{}s on the basis of their distance in the embedding space. Each cluster represents a specific entity or relation. Based on certain relevant heuristics, we assign a representative to each NP and \RP{} cluster.
\end{enumerate}

Details of different steps of \method{} are described next.


\section{\stepSide}
\label{sec:side-info}

Noun and relation phrases in Open KBs often have relevant side information in the form of useful context in the documents from which the triples were extracted. Sometimes, such information may also be present in other related KBs. Previous Open KB canonicalization methods \cite{Galarraga:2014:COK:2661829.2662073} ignored such available side information and performed canonicalization in isolation focusing only on the Open KB triples. 
\method{} attempts to exploit such side information to further improve the performance on this problem. In {\method}, we make use of five types of NP side information to get equivalence relations of the form $e_1 \equiv e_2$ between two entities $e_1$ and $e_2$. Similarly, \RP{} side information is used to derive relation equivalence, $r_1 \equiv r_2$. All equivalences are used as soft constraints in later steps of CESI (details in Section \ref{sec:embedding}).

\subsection{Noun Phrase side Information}
\label{subsec:np-side}

In the present version of {\method}, we make use of the following five types of NP side information:
\begin{enumerate}
	
	\item \textbf{Entity Linking}:  
	Given unstructured text, entity linking algorithms identify entity mentions and link them to Ontological KBs such as  Wikipedia, Freebase etc. We make use of Stanford CoreNLP entity linker which is based on \cite{SPITKOVSKY12.266} for getting NP to Wikipedia entity linking. Roughly, in about 30\% cases, we get this information for NPs. If two NPs are linked to the same Wikipedia entity, we assume them to be equivalent as per this information. For example, \textit{US} and \textit{America} can get linked to the same Wikipedia entity \textit{United\_States}.
	
	\item \textbf{PPDB Information}: 
	We make use of PPDB 2.0 \cite{DBLP:conf/acl/PavlickRGDC15}, a large collection of paraphrases in English, for identifying equivalence relation among NPs. We first extracted high confidence paraphrases from the dataset while removing duplicates. Then, using union-find, we clustered all the equivalent phrases and randomly assigned a representative to each cluster. Using an index created over the obtained clusters, we find cluster representative for each NP. If two NPs have the same cluster representative then they are considered to be equivalent. NPs not present in the dataset are skipped. This information helps us identifying equivalence between NPs such as \textit{management} and \textit{administration}. 
	
	\item \textbf{WordNet with Word-sense Disambiguation}:
	Using word-sense disambiguation \cite{Banerjee2002} with Wordnet \cite{Miller:1995:WLD:219717.219748}, we identify possible synsets for a given NP. If two NPs share a common synset, then they are marked as similar as per this side information. For example, \textit{picture} and \textit{image} can get linked to the same synset \textit{visualize.v.01}.
	
	\item \textbf{IDF Token Overlap}: 
	NPs sharing infrequent terms give a strong indication of them referring to the same entity. For example, it is very likely for \textit{Warren Buffett} and \textit{Buffett} to refer to the same person. In \cite{Galarraga:2014:COK:2661829.2662073}, IDF token overlap was found to be the most effective feature for canonicalization. We assign a score for every pair of NPs based on the standard IDF formula:
		$$score_{{idf}}(n, n') = \dfrac{\sum_{x \in w(n) \cap w(n')}{\log{(1+f(x))^{-1}}}}{\sum_{x \in w(n) \cup w(n')}{\log{(1+f(x))^{-1}}}}$$
		
	Here, $w(\cdot)$ for a given NP returns the set of its terms, excluding stop words. $f(\cdot)$ returns the document frequency for a token.  
	
	\item \textbf{Morph Normalization}: 
	We make use of multiple morphological normalization operations like tense removal, pluralization, capitalization and others as used in \cite{Fader:2011:IRO:2145432.2145596} for finding out equivalent NPs. We show in Section \ref{sec:ablation} that this information helps in improving performance. 
	
\end{enumerate}

\subsection{Relation Phrase Side Information}
Similar to noun phrases, we make use of PPDB and WordNet side information for \RP{} canonicalization as well. Apart from these, we use the following two additional types of side information involving relation phrases.

\begin{enumerate}
	\item \textbf{AMIE Information}: 
	AMIE algorithm \cite{Galarraga:2013:AAR:2488388.2488425} tries to learn implication rules between two relations $r$ and $r'$ of the form $r \Rightarrow r'$. These rules are detected based on statistical rule mining, for more details refer \cite{Galarraga:2014:COK:2661829.2662073}. It declares two relations $r$ and $r'$ to be equivalent if both $r \Rightarrow r'$ and  $r' \Rightarrow r$ satisfy support and confidence thresholds. AMIE accepts a semi-canonicalized KB as input, i.e., a KB where NPs are already canonicalized. Since this is not the case with Open KBs, we first canonicalized NPs morphologically and then applied AMIE over the NP-canonicalized KB. We chose morphological normalization for this step as such normalization is available for all NPs, and also because we found this side information to be quite effective in large Open KBs. 

	\item \textbf{KBP Information}: 
	Given unstructured text, Knowledge Base Population (KBP) systems detect relations between entities and link them to relations in standard KBs. For example, \textit{``Obama was born in Honolulu"} contains \textit{``was born in"} relation between \textit{Obama} and \textit{Honolulu}, which can be linked to \textit{per:city\_of\_birth} relation in KBs. In {\method}, we use Stanford KBP \cite{Surdeanu:2012:MML:2390948.2391003} to categorize relations. If two relations fall in the same category, then they are considered equivalent as per this information.
	
\end{enumerate}

The given list can be further extended based on the availability of other side information. For the experiments in this paper, we have used the above mentioned NP and \RP{} side information. Some of the equivalences derived from different side information might be erroneous, therefore, instead of using them as hard constraints, we try to use them as supplementary information as described in the next section. Even though side information might be available only for a small fraction of NPs and \RP{}s, the hypothesis is that it will result in better overall canonicalization. We find this to be true, as shown in Section \ref{sec:results}.

\section{\stepEmbed}
\label{sec:embedding}

For learning embeddings of NPs and \RP{}s in a given Open KB, {\method} optimizes HolE's \cite{Nickel:2016:HEK:3016100.3016172} objective function along with terms for penalizing violation of equivalence conditions from the NP and \RP{} side information. Since the conditions from side information might be spurious, a factor ($\lambda_{\text{ent}/\text{rel},\theta}$) is multiplied with each term, which acts as a hyper-parameter and is tuned on a held out validation set. We also keep a constant ($\lambda_{str}$) with HolE objective function, to make selective use of structural information from KB for canonicalization. We choose HolE because it is one of the best performing KB embeddings techniques for tasks like link prediction in knowledge graphs. Since KBs store only true triples, we generate negative examples using local closed world heuristic \cite{Dong:2014:KVW:2623330.2623623}. To keep the rank of true triples higher than the non-existing ones, we use pairwise ranking loss function. The final objective function is described below.

\begin{equation*}
\begin{split}
\min_{\Theta} 
    & \hspace{2 mm} \lambda_{str} \sum_{i \in D_+} \sum_{j \in D_-}{ \text{max} (0, \gamma + \sigma(\eta_j) - \sigma(\eta_i))} \\    
    & +  \sum_{\theta \in \mathscr{C}_{\text{ent}}} 
    	 \dfrac{\sideConst{ent}{\theta}} {|\sidePairs{ent}{\theta}|} 
    	 \sum_{v, v^{'} \in \sidePairs{ent}{\theta}}{\|e_v - e_{v^{'}}\|^2} \\
    & +  \sum_{\phi \in \mathscr{C}_{\text{rel}}} 
    	 \dfrac{\sideConst{rel}{\phi}} {|\sidePairs{rel}{\phi}|} 
    	 \sum_{u, u^{'} \in \sidePairs{rel}{\phi}}{\|r_u - r_{u^{'}}\|^2} \\
    & + \lambda_{\text{reg}} \left( \sum_{v \in V} \|e_v\|^2 + \sum_{r \in R} \|e_r\|^2 \right).
\end{split}
\end{equation*}

The objective function, consists of three main terms, along with one term for regularization. Optimization parameter, $\Theta = \{e_v\}_{v \in V} \cup \{r_u\}_{u \in R}$, is the set of all NP ($e_v$) and \RP{} ($r_u$) $d$-dimensional embeddings, where, $V$ and $R$ denote the set of all NPs and \RP{}s in the input. In the first term, $D_+, D_-$ specify the set of positive and negative examples and $\gamma > 0$ refers to the width of the margin \cite{NIPS2013_5071}. Further, $\sigma(\cdot)$ denotes the logistic function and for a triple $t_i$ $(s,p,o)$, $\eta_i = r_p^T(e_s \star e_o)$, where $\star : R^d \times R^d \rightarrow R^d$ is the circular correlation operator defined as follows.
\begin{align*}
[a \star b]_{k} = \sum_{i=0}^{d-1} a_i b_{(k+i) \text{ mod} \hspace{1 mm} d}.
\end{align*}
The first index of $(a \star b)$ measures the similarity between $a$ and $b$, while other indices capture the interaction of features from $a$ and $b$, in a particular order. Please refer to \cite{Nickel:2016:HEK:3016100.3016172} for more details.

In the second and third terms, $\mathscr{C}_{\text{ent}}$ and $\mathscr{C}_{\text{rel}}$ are the collection of all types of NP and relation side information available from the previous step (Section \ref{sec:side-info}), i.e., $\mathscr{C}_{\text{ent}} = \{ \text{Entity Linking, PPDB, ..} \}$ and $\mathscr{C}_{\text{rel}} = \{ \text{AMIE, KBP, ..} \}$. Further, $\sideConst{ent}{\theta}$ and $\sideConst{rel}{\phi}$ denote the constants associated with entity and relation side information. Their value is tuned using grid search on a held out validation set. The set of all equivalence conditions from a particular side information is denoted by $\sidePairs{ent}{\theta}$ and $\sidePairs{rel}{\phi}$. 
The rationale behind putting these terms is to allow inclusion of side information while learning embeddings, by enforcing two NPs or relations close together if they are equivalent as per the available side information. Since the side information is available for a fraction of NPs and \RP{}s in the input, including these terms in the objective does not slow down the training of embeddings significantly.

The last term adds L2 regularization on the embeddings. All embeddings are initialized by averaging GloVe vectors \cite{pennington2014glove}. We use mini-batch gradient descent for optimization.

\section{\stepCluster}
\label{sec:clustering}

{\method} clusters NPs and \RP{}s by performing Hierarchical Agglomerative Clustering (HAC) using cosine similarity over the embeddings learned in the previous step (Section \ref{sec:embedding}). HAC was preferred over other clustering methods because the number of clusters are not known beforehand. Complete linkage criterion is used for calculating the similarity between intermediate clusters as it gives smaller sized clusters, compared to single and average linkage criterion. This is more reasonable for canonicalization problem, where cluster sizes are expected to be small. The threshold value for HAC was chosen based on held out validation dataset.

%

The time complexity of HAC with complete linkage criterion is $O(n^2)$ \cite{defays1977efficient}. For scaling up {\method} to large knowledge graphs, one may go for modern variants of approximate Hierarchical clustering algorithms \cite{Kobren:2017:HAE:3097983.3098079} at the cost of some loss in performance. 

Finally, we decide a representative for each NP and \RP{} cluster. For each cluster, we compute a mean of all elements' embeddings weighted by the frequency of occurrence of each element in the input. NP or \RP{} which lies closest to the weighted cluster mean is chosen as the representative of the cluster.

\section{Experimental Setup}
\label{sec:experiments}
\vspace{-1 mm}

\subsection{Datasets}
\label{sec:datasets}

\begin{table}[t]
	\begin{tabular}{ccccc}
		\toprule
		Datasets 	& \# Gold & \#NPs 	& \#Relations 	& \#Triples \\
		& Entities & & & \\ 
		\midrule
		Base 		& 150 		& 290  & 3K 		& 9K  \\
		Ambiguous	& 446 		& 717  & 11K		& 37K \\
		\myData 	& 7.5K 		& 15.5K & 22K 		& 45K \\
		\bottomrule
		\addlinespace
	\end{tabular}
	\caption{\label{tb:datasets}Details of datasets used. \myData{} is the new dataset we propose in this paper. Please see \refsec{sec:datasets} for details.}
\end{table}


Statistics of the three datasets used in the experiments of this paper are summarized in \reftbl{tb:datasets}. We present below brief summary of each dataset.
\begin{enumerate}
	\item \textbf{Base and Ambiguous Datasets:} We obtained the Base and Ambiguous datasets from the authors of \cite{Galarraga:2014:COK:2661829.2662073}. Base dataset was created by collecting triples containing 150 sampled Freebase entities that appear with at least two aliases in ReVerb Open KB. The same dataset was further enriched with mentions of homonym entities to create the  Ambiguous dataset. Please see \cite{Galarraga:2014:COK:2661829.2662073} for more details.

	\item \textbf{\myData{}:} This is the new Open KB canonicalization dataset we propose in this paper.  
	\myData{} is a significantly extended version of the Ambiguous dataset, containing more than 20x NPs. \myData{} is constructed by intersecting information from the following three sources: ReVerb Open KB \cite{Fader:2011:IRO:2145432.2145596}, Freebase entity linking information from \cite{gabrilovich2013facc1}, and Clueweb09 corpus \cite{callan2009clueweb09}. 
	Firstly, for every triple in ReVerb, we extracted the source text from Clueweb09 corpus from which the triple was generated. In this process, we rejected  triples for which we could not find any source text. Then, based on the entity linking information from \cite{gabrilovich2013facc1}, we linked all subjects and objects of triples to their corresponding Freebase entities. If we could not find high confidence linking information for both subject and object in a triple, then it was rejected. Further, following the dataset construction procedure adopted by \cite{Galarraga:2014:COK:2661829.2662073}, we selected triples associated with all Freebase entities with at least two aliases occurring as subject in our dataset. Through these steps, we obtained 45K high-quality triples which we used for evaluation. We call this resulting dataset {\myData}.
	
	In contrast to Base and Ambiguous datasets, the number of entities, NPs and \RP{}s in \myData{} are significantly larger. 
	Please see \reftbl{tb:datasets} for a detailed comparison. This better mimics real-world KBs which tend to be sparse with very few edges per entity, as also observed by \cite{NIPS2013_5071}.


\end{enumerate}

For getting test and validation set for each dataset, we randomly sampled 20\% Freebase entities and called all the triples associated with them as validation set and rest was used as the test set. 

\subsection{Evaluation Metrics}
\label{sec:metrics}

Following \cite{Galarraga:2014:COK:2661829.2662073}, we use macro-, micro- and pairwise metrics for evaluating Open KB canonicalization methods. 
We briefly describe below these metrics for completeness. In all cases, $C$ denotes the clusters produced by the  algorithm to be evaluated, and $E$ denotes the gold standard clusters. In all cases, F1 measure is given as the harmonic mean of precision and recall.

\textbf{Macro:} Macro precision ($P_{\mathrm{macro}}$) is defined as the fraction of pure clusters in $C$,  i.e., clusters in which all the NPs (or relations) are linked to the same gold entity (or relation). Macro recall ($R_{\mathrm{macro}}$) is calculated like macro precision but with the roles of $E$ and $C$ interchanged.
\begin{eqnarray*}
	P_{\mathrm{macro}}(C, E) &=& \dfrac{|\{c \in C:\exists e \in E : e \supseteq c\}|}{|C|} \\
	R_{\mathrm{macro}}(C, E) &=& P_{\mathrm{macro}}(E, C)
\end{eqnarray*}
\textbf{Micro:} Micro precision ($P_{\mathrm{micro}}$) is defined as the purity of $C$ clusters \cite{Manning:2008:IIR:1394399} based on the assumption that the most frequent gold entity (or relation) in a cluster is correct. Micro recall ($R_{\mathrm{micro}}$) is defined similarly as macro recall.
\begin{eqnarray*} 
	P_{\mathrm{micro}}(C, E) &=& \dfrac{1}{N} \sum_{c \in C} \max_{e \in E} |c \cap e| \\
	R_{\mathrm{micro}}(C, E) &=& P_{\mathrm{micro}}(E, C)
\end{eqnarray*}
\textbf{Pairwise:} Pairwise precision ($P_{\mathrm{pair}}$) is measured as the ratio of the number of hits in $C$ to the total possible pairs in $C$. 
Whereas, pairwise recall ($R_{\mathrm{pair}}$) is the ratio of number of hits in $C$ to all possible pairs in $E$. A pair of elements in a cluster in $C$ produce a hit if they both refer to the same gold entity (or relation).
\begin{eqnarray*}
	P_{\mathrm{pair}}(C, E) &=& \dfrac{\sum_{c \in C}{|\{ (v,v') \in e, \exists e \in E, \forall (v,v') \in c\}|}} {\sum_{c \in C}{\Comb{|c|}{2}}} \\ 
	R_{\mathrm{pair}}(C, E) &=& \dfrac{\sum_{c \in C}{|\{ (v,v') \in e, \exists e \in E, \forall (v,v') \in c\}|}} {\sum_{e \in E}{\Comb{|e|}{2}}} \\
\end{eqnarray*}


\begin{figure}[t]
	\includegraphics[width=\columnwidth]{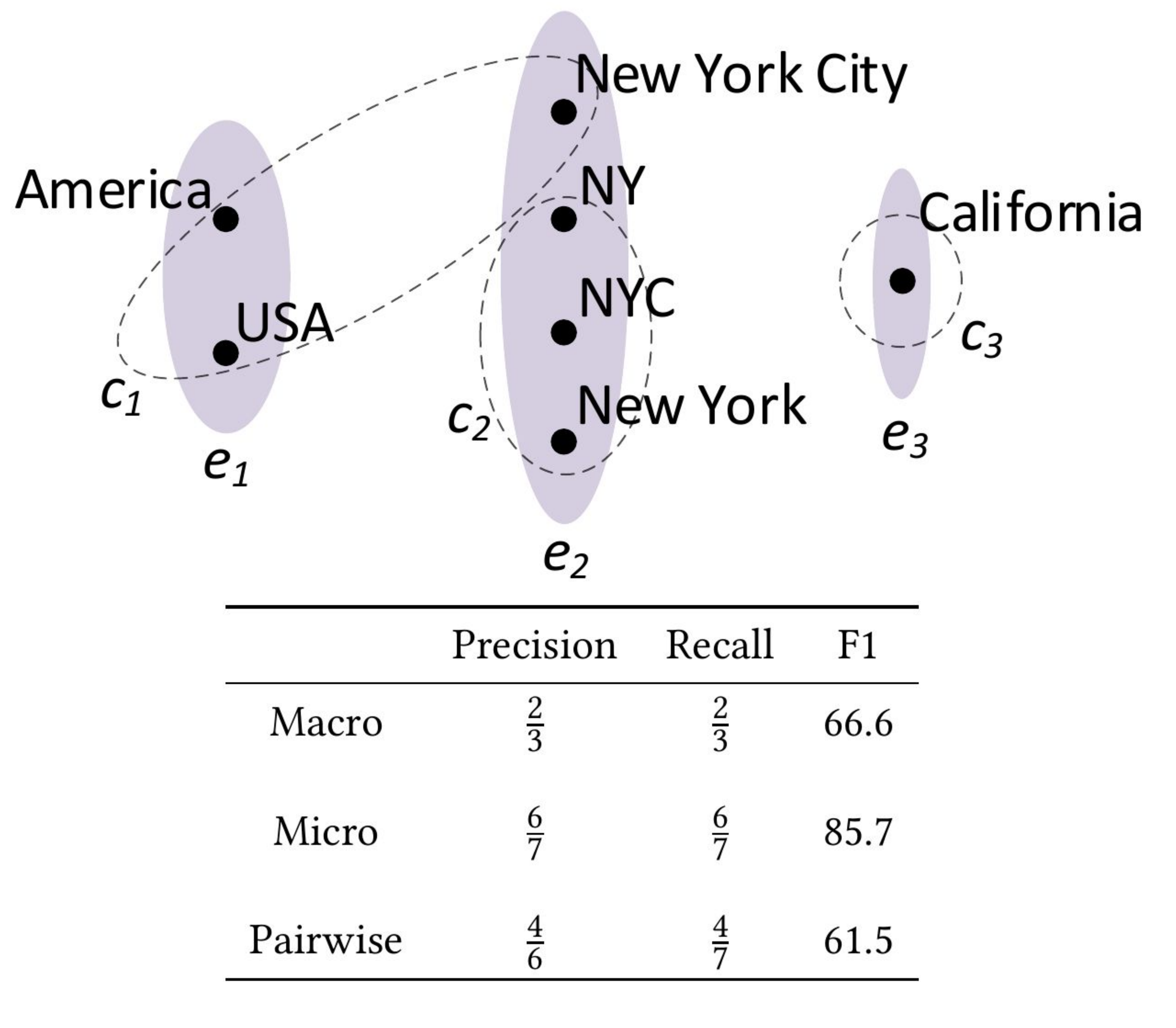}
	\caption{\label{tb:metric-example}
	Top: Illustrative example for different evaluation metrics. $e_i$ denotes actual clusters, whereas $c_i$ denotes predicted clusters. Bottom: Metric results for the above example. Please see \refsec{sec:metrics} for details.}
\end{figure}

Let us illustrate these metrics through a concrete NP canonicalization example shown in Figure \ref{tb:metric-example}. In this Figure, we can see that only $c_2$ and $c_3$ clusters in C are pure because they contain mentions of only one entity, and hence, $P_{\mathrm{macro}} = \frac{2}{3}$. On the other hand, we have $e_1$ and $e_3$ as pure clusters if we interchange the roles of $E$ and $C$. So, $R_{\mathrm{macro}} = \frac{2}{3}$ in this case. For micro precision, we can see that \textit{America}, \textit{New York}, and \textit{California} are the most frequent gold entities in $C$ clusters. Hence, $P_{\mathrm{micro}} = \frac{6}{7}$. Similarly, $R_{\mathrm{micro}} = \frac{6}{7}$ in this case.
For pairwise analysis, we need to first calculate the number of hits in $C$. In $c_1$ we have 3 possible pairs out of which only 1, (\textit{America, USA}) is a hit as they belong to same gold cluster $e_1$. Similarly, we have 3 hits in $c_2$ and 0 hits in $c_3$. Hence, $P_{\mathrm{pair}} = \frac{4}{6}$.  To compute $R_{\mathrm{pair}}$, we need total number of pairwise decisions in $E$, which is $1 + 6 + 0$ , thus, $R_{\mathrm{pair}} = \frac{4}{7}$. All the results are summarized in \reftbl{tb:metric-example}.

For evaluating NP canonicalization, we use Macro, Micro and Pairwise F1 score. However, in the case of relations, where gold labels are not available, we use macro, micro and pairwise precision values based on the scores given by human judges. 

\begin{table*}[t]
\begin{small}
	\begin{tabular}{lccc|ccc|ccc|c}
		\toprule
		Method & \multicolumn{3}{c}{Base Dataset} & \multicolumn{3}{c}{Ambiguous Dataset} & \multicolumn{3}{c}{\myData} & \\ 
		\cmidrule(r){2-4} \cmidrule(r){5-7} \cmidrule(r){8-10} \cmidrule(r){11-11} 
		& Macro & Micro & Pair. & Macro & Micro & Pair. & Macro & Micro & Pair. & Row Average\\
		\midrule
		Morph Norm	& 58.3 & 88.3 & 83.5 & 49.1 & 57.2 & 70.9 & 1.4  & 77.7 & 75.1 & 62.3\\
		PPDB       	& 42.4 & 46.9 & 32.2 & 37.3 & 60.2 & 69.3 & 46.0 & 45.4 & 64.2 & 49.3\\
		EntLinker   	& 54.9 & 65.1 & 75.2 & 49.7 & 83.2 & 68.8 & 62.8 & 81.8 & 80.4 & 69.1\\
		\gstrsim{} & 88.2 & 96.5 & 97.7 & 66.6 & 85.3 & 82.2 & 69.9 & 51.7 & 0.5  & 70.9\\
		\gidf{} & 94.8 & 97.9 & 98.3 & 67.9 & 82.9 & 79.3 & 71.6 & 50.8 & 0.5  & 71.5\\
		\gattr{}	& 76.1 & 51.4 & 18.1 & \textbf{82.9} & 27.7 & 8.4 & \textbf{75.1} & 20.1 & 0.2  & 40.0\\
		GloVe 		& 95.7 & 97.2 & 91.1 & 65.9 & 89.9 & 90.1 & 56.5 & 82.9 & 75.3 & 82.7\\
		HolE (Random)		& 69.5 & 91.3 & 86.6 & 53.3 & 85.0 & 75.1 & 5.4  & 74.6 & 50.9 & 65.7\\
		HolE (GloVe)    & 75.2 & 93.6 & 89.3 & 53.9 & 85.4 & 76.7 & 33.5 & 75.8 & 51.0 & 70.4\\
		\method 	& \textbf{98.2} & \textbf{99.8} & \textbf{99.9} & 66.2 & \textbf{92.4} & \textbf{91.9} & 62.7 & \textbf{84.4} & \textbf{81.9} & \textbf{86.3}\\
		\bottomrule
		\addlinespace
	\end{tabular}
	\caption{\label{tb:np_canonicalization}NP Canonicalization Results. \method{} outperforms all other methods across datasets (Best in 7 out of 9 cases. \refsec{sec:np_results})}
\end{small}
\end{table*}

\subsection{Methods Compared}
\subsubsection{\bf Noun Phrase Canonicalization}  
For \NP{} canonicalization, {\method} has been compared against the following methods:

\begin{itemize}
	\item \textbf{Morphological Normalization:} As used in \cite{Fader:2011:IRO:2145432.2145596}, this involves applying simple normalization operations like removing tense, pluralization, capitalization etc. over NPs and \RP{}s.
	\item {\bf Paraphrase Database (PPDB):} Using PPDB 2.0 \cite{DBLP:conf/acl/PavlickRGDC15}, we clustered two NPs together if they happened to share a common paraphrase. NPs which could not be found in PPDB are put into singleton clusters.
	\item {\bf Entity Linking}: Since the problem of \NP{} canonicalization is closely related to entity linking, we compare our method against Stanford CoreNLP Entity Linker \cite{SPITKOVSKY12.266}. Two NPs linked to the same entity are clustered together.
	\item \textbf{\gidf{} \cite{Galarraga:2014:COK:2661829.2662073}}: IDF Token Overlap was the best performing method proposed in \cite{Galarraga:2014:COK:2661829.2662073} for NP canonicalization. In this method, IDF token similarity is defined between two NPs as in  \refsec{subsec:np-side}, and HAC is used to cluster the mentions.
	\item \textbf{\gstrsim{} \cite{Galarraga:2014:COK:2661829.2662073}}: This method is similar to Galarraga-IDF, but with similarity metric being the Jaro-Winkler \cite{winkler1999state} string similarity measure.
	\item \textbf{\gattr{} \cite{Galarraga:2014:COK:2661829.2662073}}: Again, this method is similar to the Galarraga-IDF, except that Attribute Overlap is used as the similarity metric between two NPs in this case. Attribute for a NP $n$, is defined as the set of relation-NP pairs which co-occur with $n$ in the input triples. Attribute overlap similarity between two NPs, is defined as the Jaccard coefficient of the set of attributes:
	$$ f_{\text{attr}}(n,n') = \dfrac{|A \cap A'|}{|A \cup A'|}$$
	where, $A$ and $A'$ denote the set of attributes associated with $n$ and $n'$.
	
	Since canonicalization methods using above similarity measures were found to be most effective in \cite{Galarraga:2014:COK:2661829.2662073}, even outperforming Machine Learning-based alternatives, we consider these three baselines as representatives of state-of-the-art in Open KB canonicalization. 
	
	\item {\bf GloVe}: 
	In this scheme, each NP and \RP{} is represented by a 300 dimensional GloVe embedding \cite{pennington2014glove} trained on Wikipedia 2014 and Gigaword 5 \cite{parker2011english} datasets with 400k vocabulary size. Word vectors were averaged together to get embeddings for multi-word phrases. These GloVE embeddings were then clustered for final canonicalization.
	
	\item {\bf HolE}: In this method, embeddings of NPs and \RP{}s in an Open KB are obtained by applying HolE \cite{Nickel:2016:HEK:3016100.3016172} over the Open KB. These embeddings are then clustered to obtain the final canonicalized groupings. Based on the initialization of embeddings, we differentiate between \textbf{HolE(Random)} and \textbf{HolE(GloVe)}. 
	\item {\bf CESI}: This is the method proposed in this paper, please see \refsec{sec:approach} for more details. 

\end{itemize}

{\bf Hyper-parameters}: Following \cite{Galarraga:2014:COK:2661829.2662073}, we used Hierarchical Agglomerative Clustering (HAC) as the default clustering method across all methods (wherever necessary). For all methods, grid search over the hyperparameter space was performed, and results for the best performing setting are reported. This process was repeated for each dataset.

\subsubsection{\bf Relation Phrase Canonicalization} AMIE \cite{Galarraga:2013:AAR:2488388.2488425} was found to be effective for \RP{} canonicalization in \cite{Galarraga:2014:COK:2661829.2662073}. We thus consider AMIE\footnote{We use support and confidence values of 2 and 0.2 for all the experiments in this paper.} as the state-of-the-art baseline for \RP{} canonicalization and compare against \method{}. We note that AMIE requires NPs of the input Open KB to be already canonicalized. In all our evaluation datasets, we already have \emph{gold} NP canonicalization available. We provide this gold NP canonicalization information as input to AMIE. Please note that \method{} doesn't require such pre-canonicalized NP as input, as it performs \emph{joint} NP and \RP{} canonicalization. Moreover, providing gold NP canonicalization information to AMIE puts \method{} at a disadvantage. We decided to pursue this choice anyways in the interest of stricter evaluation. However, in spite of starting from this disadvantageous position, \method{} significantly outperforms AMIE in \RP{} canonicalization, as we will see in \refsec{sec:rel-eval}.


For evaluating performance of both  algorithms, we randomly sampled 25 non-singleton relation clusters for each of the three datasets and gave them to five different human evaluators\footnote{Authors did not participate in this evaluation.} for assigning scores to each cluster. The setting was kept blind, i.e., identity of the algorithm producing a cluster was not known to the evaluators. Based on the average of evaluation scores, precision values were calculated. Only non-singleton clusters were sampled, as singleton clusters will always give a precision of one.



\section{Results}
\label{sec:results}

In this section, we evaluate the following questions.
\begin{itemize}
	\item[Q1.] Is \method{} effective in Open KB canonicalization? (\refsec{sec:overall-perf})
	\item[Q2.] What is the effect of side information in \method{}'s performance? (\refsec{sec:ablation})
	\item[Q3.] Does addition of entity linking side information degrade \method{}'s ability to canonicalize unlinked NPs (i.e., NPs missed by the entity linker)? (\refsec{sec:unlinked-eval})
\end{itemize}

Finally, in \refsec{sec:qualitative}, we present qualitative examples and discussions.

\subsection{Evaluating Effectiveness of \method{} in Open KB Canonicalization}
\label{sec:overall-perf}

\subsubsection{\bf Noun Phrase Canonicalization}

\label{sec:np_results}
Results for \NP{} canonicalization are summarized in \reftbl{tb:np_canonicalization}. Overall, we find that \method{} performs well consistently across the datasets. 
Morphological Normalization failed to give competitive performance in presence of homonymy.  
PPDB, in spite of being a vast reservoir of paraphrases, lacks information about real-world entities like people, places etc. Therefore, its performance remained weak throughout all datasets. Entity linking methods 
make use of contextual information from source text of each triple to link a NP to a KB entity. But their performance is limited because they are restricted by the entities in KB.
String similarity also gave decent performance in most cases but since they solely rely on surface form of NPs, they are bound to fail with NPs having dissimilar mentions.

Methods such as \gidf{}, \gstrsim{}, and \gattr{} performed poorly on \myData{}. Although, their performance is considerably better on the other two datasets. This is because of the fact that in contrast to Base and Ambiguous datasets, {\myData} has considerably large number of entities and comparatively fewer triples (\reftbl{tb:datasets}). \gidf{} token overlap is more likely to put two NPs together if they share an uncommon token, i.e., one with high IDF value. Hence, accuracy of the method relies heavily on the quality of document frequency estimates which may be quite misleading when we have smaller number of triples. Similar is the case with \gattr{} which decides similarity of NPs based on the set of shared attributes. Since, attributes for a NP is defined as a set of relation-NP pairs occurring with it across all triples, sparse data also results in poor performance for this method. 

GloVe captures semantics of NPs and unlike string similarity it doesn't rely on the surface form of NPs. Therefore, its performance has been substantial across all the datasets. HolE captures structural information from the given triples and uses it for learning embeddings. Through our experiments, we can see that solely structural information from KB is quite effective for NP canonicalization. {\method} performs the best across the datasets in 7 out of the 9 settings, as it incorporates the strength of all the listed methods. The superior performance of {\method} compared to HolE clearly indicates that the side information is indeed helpful for canonicalization task. Results of GloVe, HolE and {\method} suggest that embeddings based method are much more  effective for Open KB canonicalization.

\begin{table}[t]
	\begin{tabular}{ccccc}
		\toprule
		& Macro & Micro & Pairwise & Induced \\
		& Precision & Precision & Precision &  Relation \\
		&  &  &  &  Clusters \\
		\midrule		
		\addlinespace
			\multicolumn{5}{c}{\textbf{Base Dataset}} \\
			AMIE    & 42.8 & 63.6 & 43.0 & 7 \\
			\method & \textbf{88.0} & \textbf{93.1} & \textbf{88.1} & \textbf{210} \\	
		\addlinespace
		\hline
		\addlinespace
			\multicolumn{5}{c}{\textbf{Ambiguous Dataset}} \\ 
			AMIE    & 55.8 & 64.6 & 23.4 & 46 \\
			\method & \textbf{76.0} & \textbf{91.9} & \textbf{80.9} & \textbf{952}\\					
		\addlinespace
		\hline
		\addlinespace 
			\multicolumn{5}{c}{\textbf{\myData}} \\ 
			AMIE 	& 69.3 & 84.2 & 66.2 & 51 \\
			\method & \textbf{77.3} &\textbf{87.8} & \textbf{72.6} & \textbf{2116} \\
		\bottomrule
		\addlinespace
	\end{tabular}
	\caption{\label{tb:rel_canonicalization}Relation canonicalization results. Compared to AMIE, \method{} canonicalizes more number of \RP{}s at higher precision. Please see \refsec{sec:rel-eval} for details.}
\end{table}

\subsubsection{\bf Relation Phrase Canonicalization}
\label{sec:rel-eval}

Results for \RP{} canonicalization are presented in \reftbl{tb:rel_canonicalization}. For all experiments, in spite of using  quite low values for minimum support and confidence, AMIE was unable to induce any reasonable number of non-singleton clusters (e.g., only 51 clusters out of the 22K \RP{}s in the \myData{} dataset). For relation canonicalization experiments, AMIE was evaluated on gold NP canonicalized data as the algorithm requires NPs to be already canonicalized. \method{}, on the other hand, was tested on all the datasets without making use of gold NP canonicalization information. 

Based on the results in \reftbl{tb:rel_canonicalization}, it is quite evident that AMIE induces too few relation clusters to be of value in practical settings. 
On the other hand, {\method} consistently performs well across all the datasets and induces  significantly larger number of clusters. 


\subsection{Effect of Side Information in \method{}}
\label{sec:ablation}

\begin{figure}[t]
	\includegraphics[width=\columnwidth]{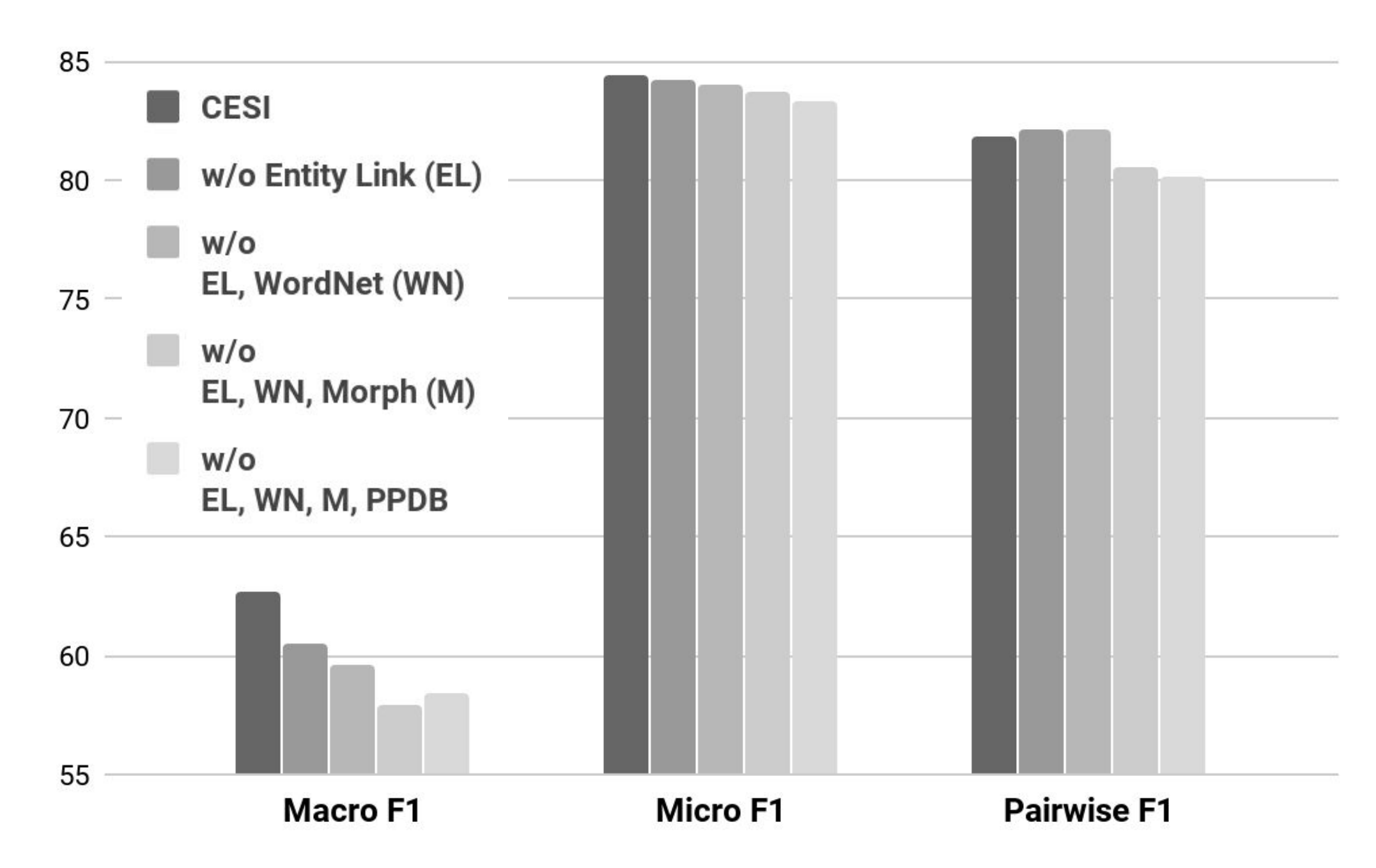}
	\caption{\label{fig:ablation}Performance comparison of various side information-ablated versions of \method{} for NP canonicalization in the \myData{} dataset. Overall, side information helps \method{} improve performance. Please see \refsec{sec:ablation} for details.}
\end{figure}

In this section, we evaluate the effect of various side information in \method{}'s performance. For this, we evaluated the performances of various  versions of \method{}, each one of them obtained by ablating increasing amounts of side information from the full \method{} model. Experimental results comparing these ablated versions on the \myData{} are presented in \reffig{fig:ablation}. From this figure, we observe that while macro performance benefits most from different forms of side information, micro and pairwise performance also show increased performance in the presence of various side information. This validates one of the central thesis of this paper: side information, along with embeddings, can result in improved Open KB canonicalization.


\begin{table}[t]
	\begin{tabular}{lccc}
		\toprule
		& Macro F1 	& Micro F1 	& Pairwise F1 \\ 
		\midrule
		\method 			& 81.7 & 87.6 & 81.5 	\\
		{\method} w/o EL	& 81.3 & 87.3 & 80.7	\\
		\bottomrule
		\addlinespace
	\end{tabular}
	\caption{\label{tb:non-linked}\method{}'s performance in canonicalizing unlinked NPs, with and without Entity Linking (EL) side information, in the \myData{} dataset. We observe that \method{} does not overfit to EL side information, and thereby helps prevent performance degradation in unlinked NP canonicalization (in fact it even helps a little). Please see \refsec{sec:unlinked-eval} for details.}
\end{table}

\subsection{Effect of Entity Linking Side Information on Unlinked NP Canonicalization}
\label{sec:unlinked-eval}

From experiments in \refsec{sec:ablation}, we find that Entity Linking (EL) side information (see \refsec{subsec:np-side}) is one of the most useful side information that \method{} exploits. However, such side information is not available in case of unlinked NPs, i.e., NPs which were not linked by the entity linker. So, this naturally raises the following question: does \method{} overfit to the EL side information and ignore the unlinked NPs, thereby resulting in poor canonicalization of such unlinked NPs?

In order to evaluate this question, we compared \method{}'s performance on unlinked NPs in the \myData{} dataset, with and without EL side information. We note that triples involving unlinked NPs constitute about 25\% of the entire dataset. Results are presented in \reftbl{tb:non-linked}. From this table, we observe that \method{} doesn't overfit to EL side information, and it selectively uses such information when appropriate (i.e., for linked NPs). Because of this robust nature, presence of EL side information in \method{} doesn't have an adverse effect on the unlinked NPs, in fact there is a small  gain in performance. 


\begin{figure}[t]
	\fbox{\includegraphics[width=\columnwidth]{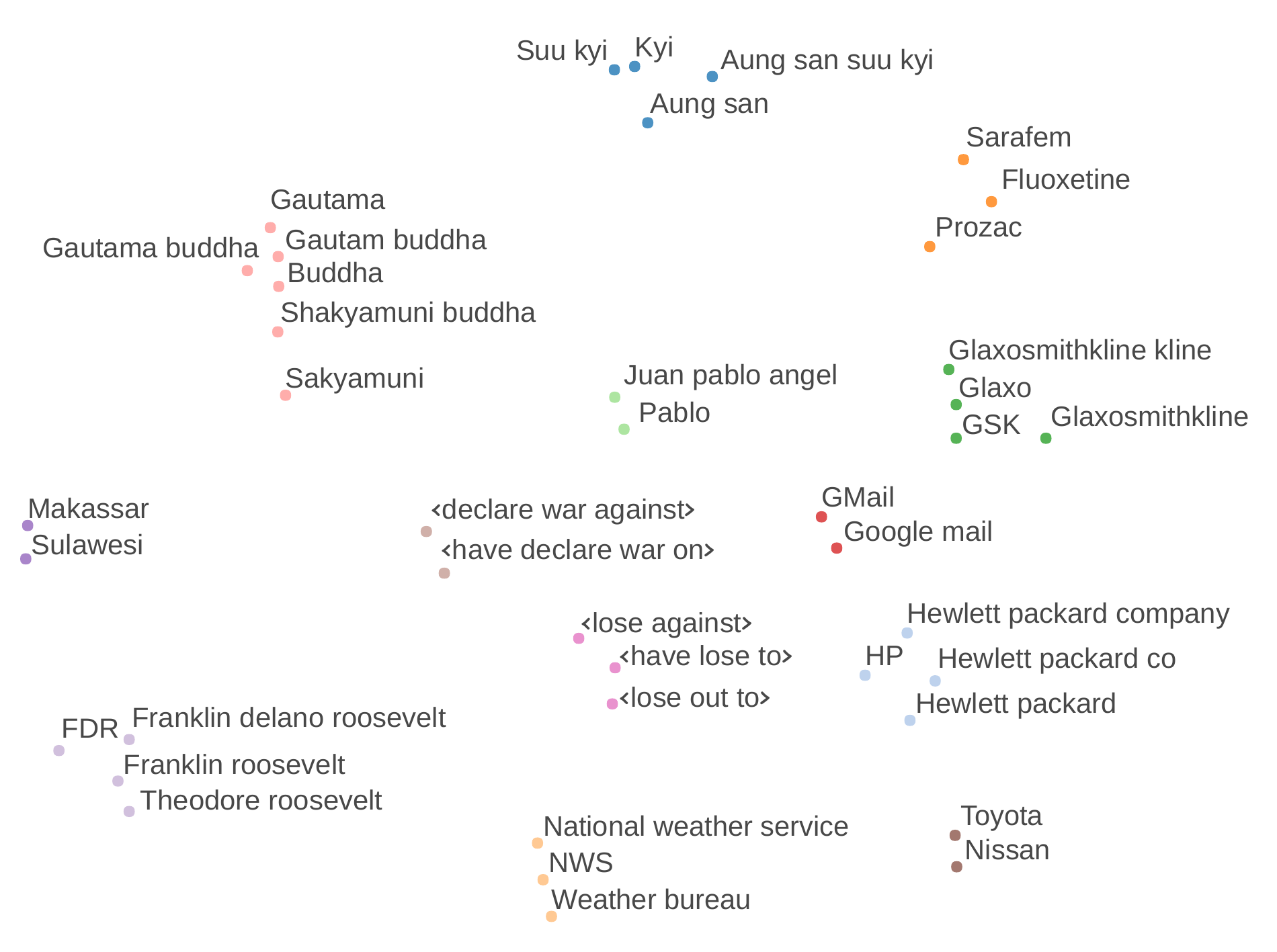}}
	\caption{\label{fig:embeddings}t-SNE visualization of NP and relation phrase (marked in '$<\cdots>$') embeddings learned by {\method} for {\myData} dataset. We observe that \method{} is able to induce non-trivial canonical clusters. 
	Please see \refsec{sec:qualitative} for details.}	
\end{figure}



\subsection{Qualitative Evaluation}
\label{sec:qualitative}

Figure \ref{fig:embeddings} shows some of the NP and relation phrase clusters detected by {\method} in {\myData} dataset. These results highlight the efficacy of algorithm in canonicalizing non-trivial NPs and relation phrases. The figure shows t-SNE \cite{maaten2008visualizing} visualization of NP and relation phrase (marked in '$<\cdots>$') embeddings for a few examples. We can see that the learned embeddings are actually able to capture equivalence of NPs and relation phrases. The algorithm is able to correctly embed \textit{Prozac}, \textit{Sarafem} and \textit{Fluoxetine} together (different names of the same drug), despite their having completely different surface forms. 

Figure \ref{fig:embeddings} also highlights the failures of {\method}. For example, \textit{Toyota} and \textit{Nissan} have been embedded together although the two being different companies. Another case is with \textit{Pablo} and \textit{Juan Pablo Angel}, which refer to different entities. The latter case can be avoided by keeping track of the source domain type information of each NP for disambiguation. In this if we know that \textit{Juan Pablo Angel} has come from \textit{SPORTS} domain, whereas \textit{Pablo} has come from a different domain then we can avoid putting them together. We tried using DMOZ \cite{OMV93V_2016} dataset, which provide mapping from URL domain to their categories, for handling such errors. But, because of poor coverage of URLs in DMOZ dataset, we couldn't get significant improvement in canonicalization results. We leave this as a future work.

\section{Conclusion}
\label{sec:conclusion}


Canonicalizing Open Knowledge Bases (KBs) is an important but underexplored problem. In this paper, we proposed \method{}, a novel method for canonicalizing Open KBs using learned embeddings and side information. 
\method{} solves a joint objective to learn noun and relation phrase embeddings, while utilizing relevant side information in a principled manner. These learned embeddings are then clustered together to obtain canonicalized noun and relation phrase clusters. In this paper, we also propose \myData{}, a new and larger dataset for Open KB canonicalization. Through extensive experiments on this and other real-world datasets, we demonstrate \method{}'s effectiveness over state-of-the-art baselines. \method{}'s source code and all data used in the paper are publicly available\footnote{\texttt{\url{https://github.com/malllabiisc/cesi}}}.


\section*{Acknowledgement}

We thank the reviewers for their constructive comments. This work is supported in part by MHRD, Govt. of India, and by gifts from Google Research and Accenture. We thank Anand Mishra and other members of MALL Lab, IISc for carefully reading drafts of this paper.

\bibliographystyle{ACM-Reference-Format}
\balance
\bibliography{references.bib}

\end{document}